\documentclass[amsmath,amssymb,superscriptaddress,nobalancelastpage,aps,prl,twocolumn]{revtex4-2}

\usepackage{graphicx}
\usepackage{varioref}
\usepackage{xr-hyper}
\usepackage{xcolor}
\usepackage{nicefrac}
\usepackage{xfrac}
\usepackage{hyperref}
\hypersetup{colorlinks,linkcolor=blue,urlcolor=blue,citecolor=blue}
\usepackage{ulem}
\usepackage{lineno}
\usepackage{amsmath} 
\usepackage{amssymb}

\newcommand{\EuLSCO}{La$_{1.675}$Eu$_{0.2}$Sr$_{0.125}$CuO$_4$}

\begin{document}

\title{Charge Order Lock-in by Electron-Phonon Coupling in La$_{1.675}$Eu$_{0.2}$Sr$_{0.125}$CuO$_4$}

\author{Qisi~Wang}
\email{qisiwang@physik.uzh.ch}
\affiliation{Physik-Institut, Universit\"{a}t Z\"{u}rich, Winterthurerstrasse 
190, CH-8057 Z\"{u}rich, Switzerland}

\author{K.~von~Arx}
\affiliation{Physik-Institut, Universit\"{a}t Z\"{u}rich, Winterthurerstrasse 
190, CH-8057 Z\"{u}rich, Switzerland}

\author{M.~Horio}
\affiliation{Physik-Institut, Universit\"{a}t Z\"{u}rich, Winterthurerstrasse 
190, CH-8057 Z\"{u}rich, Switzerland}

\author{D.~John~Mukkattukavil}
\affiliation{Department of Physics and Astronomy, Uppsala University, Box 516, SE-751 20 Uppsala, Sweden}

\author{J. K\"{u}spert}
\affiliation{Physik-Institut, Universit\"{a}t Z\"{u}rich, Winterthurerstrasse 190, CH-8057 Z\"{u}rich, Switzerland}

\author{Y.~Sassa}
\affiliation{Department of Physics, Chalmers University of Technology, SE-412 96 G\"{o}teborg, Sweden}

\author{T.~Schmitt}
\affiliation{Swiss Light Source, Photon Science Division, Paul Scherrer Institut, CH-5232 Villigen PSI, Switzerland}

\author{A.~Nag}
\affiliation{Diamond Light Source, Harwell Campus, Didcot, Oxfordshire OX11 0DE, United Kingdom}

\author{S.~Pyon}
\affiliation{Department of Advanced Materials, University of Tokyo, Kashiwa 277-8561, Japan}

\author{T.~Takayama}
\affiliation{Department of Advanced Materials, University of Tokyo, Kashiwa 277-8561, Japan}

\author{H.~Takagi}
\affiliation{Department of Advanced Materials, University of Tokyo, Kashiwa 277-8561, Japan}

\author{M.~Garcia-Fernandez}
\affiliation{Diamond Light Source, Harwell Campus, Didcot, Oxfordshire OX11 0DE, United Kingdom}

\author{Ke-Jin~Zhou}
\affiliation{Diamond Light Source, Harwell Campus, Didcot, Oxfordshire OX11 0DE, United Kingdom}

\author{J.~Chang}
\email{johan.chang@physik.uzh.ch}
\affiliation{Physik-Institut, Universit\"{a}t Z\"{u}rich, Winterthurerstrasse 190, CH-8057 Z\"{u}rich, Switzerland}

\begin{abstract}
We report an ultrahigh resolution resonant inelastic x-ray scattering (RIXS) study of the in-plane bond-stretching phonon mode in stripe-ordered cuprate La$_{1.675}$Eu$_{0.2}$Sr$_{0.125}$CuO$_4$. Phonon softening and lifetime shortening are found around the charge ordering wave vector. In addition to these self-energy effects, the electron-phonon coupling is probed by its proportionality to the RIXS cross section. We find an enhancement of the electron-phonon coupling around the charge-stripe ordering wave vector upon cooling into the low-temperature tetragonal structure phase. These results suggest that in addition to electronic correlations, electron-phonon coupling contributes significantly to the emergence of long-range charge-stripe order in cuprates.

\end{abstract}

\maketitle
The ubiquity of charge order in hole-doped cuprates has motivated the investigation of its underlying mechanism~\cite{KeimerNat2015}. This task is complicated by the fact that charge, spin and lattice degrees of freedom are intimately coupled in the presence of strong correlations. Prevalent theories suggest that charge order in underdoped cuprates develops predominantly from strong electronic interactions~\cite{TranquadaNat1995,EmeryPNAS1999,ZaanenPRB1989}. The competition between magnetic and kinetic energies drives the tendency of doped holes towards clustering into charged stripes --- often dubbed stripe order. This is in contrast to the scenario where charge-density-wave (CDW) order emerges due to momentum dependent electron-phonon coupling (EPC)~\cite{MazinPRB2008,WeberPRL2011,ZhuPNAS2015}. These two mechanisms for charge ordering are difficult to differentiate since (a) charge modulation in both cases couples to the underlying lattice and (b) it is difficult to evaluate the EPC and its reciprocal space variation. From hereon, CDW order refers broadly to a static charge modulation irrespective of the driving mechanism which is the topic of this Letter.

Phonon anomalies in vicinity to the CDW ordering wave vector suggest the importance of electron-phonon interaction~\cite{PintschoviusPSSB2005,ReznikPhysC2012}. It has, however, been difficult to extract the EPC strength let alone its momentum dependence. Inelastic neutron and x-ray scattering (INS and IXS)~\cite{PintschoviusPSSB2005,ReznikPhysC2012} or angle-resolved photoemission spectroscopy (ARPES) self-energy studies~\cite{DamascelliRMP2003,CukPSSB2005}, infer the EPC strength based on the assumption of bare phononic or electronic dispersions. Typically, the experimentally extracted self-energies based on these assumptions are not sensitive to subtle crystalline structural changes. Yet, in underdoped cuprates, charge order seems to be favored by certain crystal structures. For example, the low-temperature tetragonal (LTT) structure appears as the ideal host for CDW order~\cite{TranquadaNat1995,FujitaPRL2002, FujitaPRB2002}. This leads to the perception that the LTT phase enhances the coupling to the electronic stripes~\cite{TranquadaNat1995,KampfPRB2001}. It has, however, not been possible to experimentally establish any relation between the specific crystal structures and electron-phonon interaction.

\begin{figure*}[ht!]
 	\begin{center}
 		\includegraphics[width=0.99\textwidth]{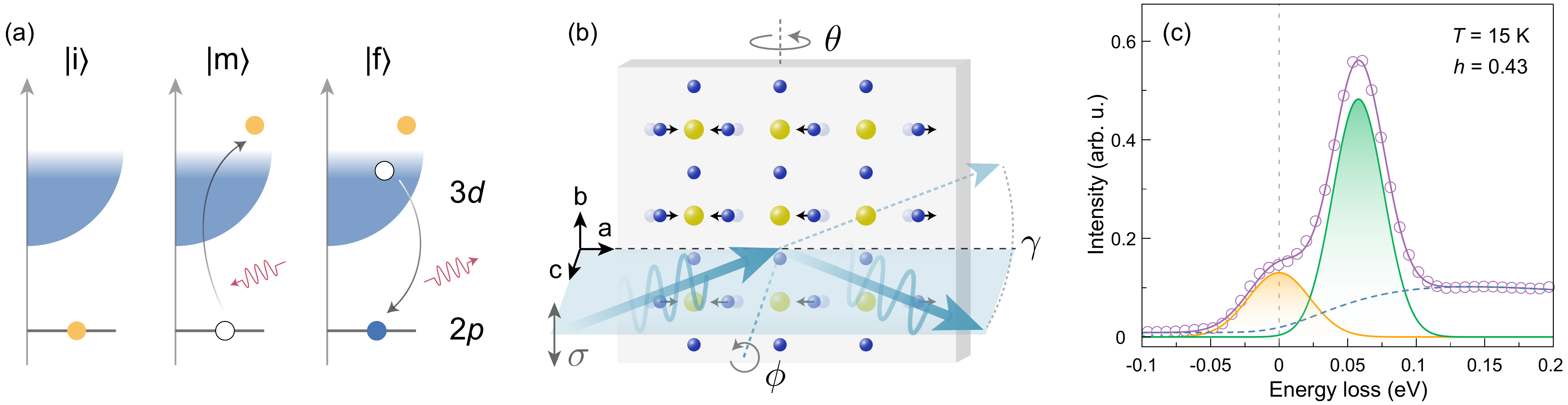}
 	\end{center}
 	\caption{Phonon excitations generated in a RIXS process. (a) Schematic of a two-step Cu $L_3$-edge RIXS process. The transition from initial  $|i\rangle$ to the intermediate state $|m\rangle$ occurs upon absorption of an incident photon, a Cu $2p$ core electron is promoted to the $3d$ valence band. While the core hole is screened, the excited electron effectively causes the vibrations of the oxygen ions. A decay process leaves behind a phonon in the final state $|f\rangle$ with probability depending on the electron-phonon coupling~\cite{AmentEPL2011}. (b) Schematic of the scattering geometry and the eigenvector for phonons with $\mathbf{Q}=(0.25,0,0)$. The yellow and blue ``balls" represent copper and oxygen ions, respectively. Black arrows on the oxygen ions indicate the vibration directions. (c) A representative RIXS spectrum collected at 15 K on \EuLSCO. The orange, green solid and blue dashed lines represent scattering from elastic, phonon and background contributions, respectively. The fitting components are described in the text.
 	}	
	 	\label{fig:fig1}
 \end{figure*}

Recent advances of the resonant inelastic x-ray scattering (RIXS) technique have enabled studies of optical phonons, charge excitations, and their Fano interference~\cite{ChaixNP2017,MiaoPRX2019,LeeNP2020,LiPNAS2020,LinPRL2020,PengPRL2020}.
The RIXS phonon scattering cross section is directly proportional to the EPC strength [see Figs.~\ref{fig:fig1}(a) and \ref{fig:fig1}(b)]~\cite{AmentRMP2011,AmentEPL2011,BraicovichPRR2020,RossiPRL2019,DevereauxPRX2016,JohnstonPRB2010}. It is therefore possible to study both the self-energy effects (excitation energy shift and lifetime) and the EPC strength for specific phonon modes.

Here we use ultrahigh energy resolution RIXS to address the electron-phonon coupling in the stripe-ordered compound \EuLSCO~(LESCO). We show how the EPC  is enhanced upon cooling from the low-temperature orthorhombic (LTO) into the low-temperature tetragonal phase that boosts the CDW order. Specifically, we study the in-plane copper-oxygen bond-stretching phonon mode as a function of temperature and momentum. A phonon softening anomaly in vicinity to the CDW ordering wave vector is found to persist beyond the LTT phase. In the same momentum range, we find that the EPC strength is enhanced upon cooling into the LTT phase. The enhanced EPC shows a strong momentum dependence peaking near the charge ordering wave vector. While the phonon anomaly corresponds to an energy  softening of less than 10$\%$, the EPC is enhanced by about 20$\%$ in the LTT phase. Our results suggest that the enhanced EPC amplifies the CDW order. Both electronic correlations and momentum dependent EPC are responsible for the charge ordering in this cuprate compound. While fluctuating stripes emerge spontaneously from strong electron interactions at high temperature, they are stabilized at low temperature via the enhanced coupling to the host lattice.

Single crystals of LESCO, grown by the floating zone method, were studied at the I21 RIXS beamline at Diamond Light Source. All measurements were performed at the Cu $L_{3}$ resonance ($\sim932.5$~eV), using grazing exit geometry with linear vertical ($\sigma$) incident light polarization. 
We define wave vector $\mathbf{Q}$ at $(q_{x},q_{y},q_{z})$ as $(h,k,\ell)=(q_{x}a/2\pi,q_{y}b/2\pi,q_{z}c/2\pi)$ in reciprocal lattice units (r.l.u.) using pseudo-tetragonal notation, with $a\approx b\approx3.79$~\AA~and $c\approx13.1$~\AA. 
Scattering angle and energy resolution (Gaussian standard deviation) were fixed to $\gamma=154^{\circ}$ and $\sigma_G=19$ meV. Magnitude and direction of the in-plane momentum transfer $\mathbf{Q}_{\parallel}=(h,h \tan \phi)$ are controlled by the light incident angle $\theta$ and sample azimuthal angle $\phi$, respectively [Fig.~\ref{fig:fig1}(b)]. RIXS intensities were normalized to the weight of the $dd$ excitations (see Supplemental Material~\cite{SM_note} for details).

\begin{figure*}[ht!]
 	\begin{center}
 		\includegraphics[width=0.99\textwidth]{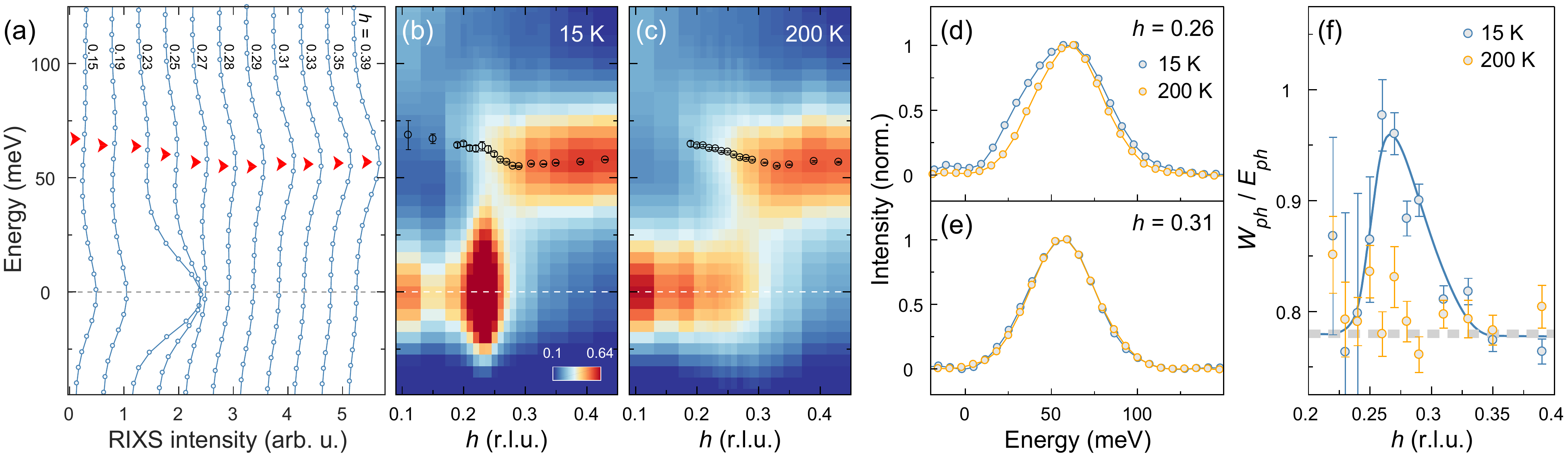}
 	\end{center}
 	\caption{
 	Bond-stretching phonon mode in \EuLSCO\ probed by RIXS. (a) Raw RIXS spectra collected at 15~K across the charge ordering wave vector. (b), (c) RIXS intensity maps as a function of energy loss and $h$ for temperatures as indicated. The red arrows in (a) and black open circles in (b) and (c) mark the phonon dispersion determined from fitting the spectra (see text). Color code indicates the RIXS intensity. To avoid the overwhelming elastic scattering at the charge ordering wave vector, we here present data taken along $\mathbf{Q}=(h,h\tan \phi)$ with $\phi=4^{\circ}$ and $0^{\circ}$ for respectively $T=15$ and 200~K. (d), (e) Elastic scattering and background subtracted spectra for $h=0.26$ and $0.31$. (f) Dimensionless parameter $W_{ph}/E_{ph}$ as a function of $h$ at 15 K and 200 K. Solid lines are guides to the eye. Error bars, throughout the Letter, reflect standard deviations of the fits described in the text.
 	}	
	 	\label{fig:fig2}
 \end{figure*}

\begin{figure*}[ht!]
 	\begin{center}
 		\includegraphics[width=0.99\textwidth]{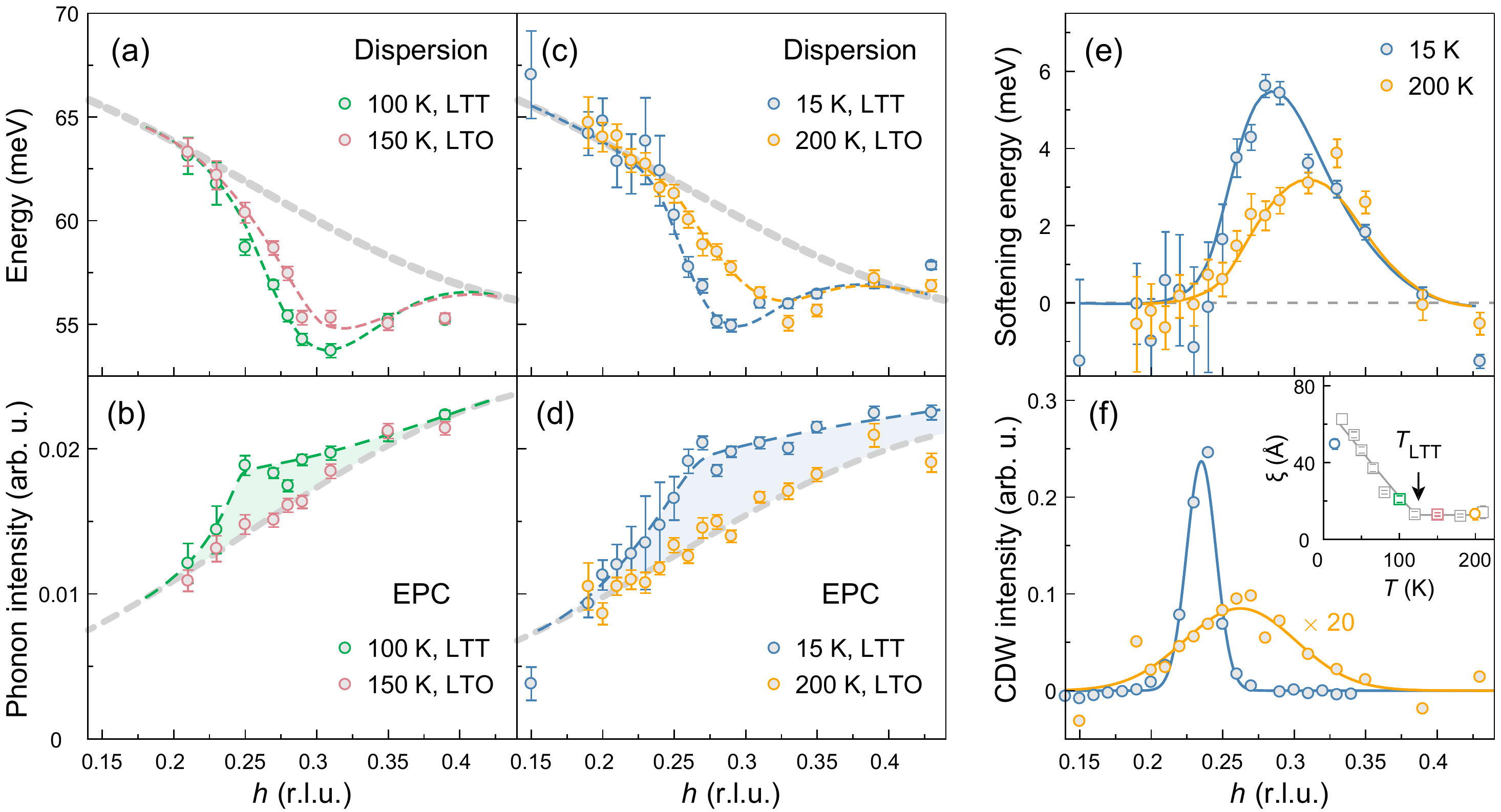}
 	\end{center}
 	\caption{
 	Temperature evolution of charge order and the associated phonon anomaly in \EuLSCO. (a), (c) Phonon dispersions and (b), (d) integrated spectral weights along the longitudinal direction at indicated temperatures. (e) Softening energy as a function of $h$ obtained from data in (c) as described in the text. (f) Charge ordering peak at 15 K and 200 K. Elastic intensities are integrated over $-45$ to $30$~meV and the 200~K data are displayed with a multiplication factor of 20. Solid lines are Gaussian fits. The inset displays the charge order correlation length $\xi$ versus temperature. Open squares denote data from Ref.~\cite{WangPRL2020}. Gray dashed lines in (a),(c) and (b),(d) are respectively the DFT bare bond-stretching phonon dispersion (normalized to match our data) and the phonon spectral weight above $T\rm_{LTT}$ fitted to a $\sin^2(\pi h)$ function. Colored lines in (a)-(e) are guides to the eye.
 	}	
	 	\label{fig:fig3}
 \end{figure*}

A RIXS spectrum obtained at $\mathbf{Q}=(0.43,0.03)$ [Fig.~\ref{fig:fig1}(c)], reveals a pronounced excitation around 60~meV that dominates over elastic scattering and the paramagnon excitations. We assign this excitation to the in-plane copper-oxygen bond-stretching phonon [see Fig.~\ref{fig:fig1}(b)], in agreement with previous INS~\cite{ReznikNat2006}, IXS~\cite{FukudaPRB2005} and RIXS studies~\cite{PengPRL2020,LinPRL2020}. To determine its dispersion, RIXS spectra are collected as a function of momentum [Figs.~\ref{fig:fig2}(a) and S2]. The resulting RIXS intensity maps are shown for low (15~K) and high (200~K) temperatures in Figs.~\ref{fig:fig2}(b) and \ref{fig:fig2}(c). Stripe order, revealed by the reflection at $\mathbf{Q}\rm_{CDW}=(\delta, 0)$ with $\delta\approx0.24$, weakens by about two orders of magnitude across this temperature range [see Fig.~\ref{fig:fig3}(f) and Ref.~\onlinecite{WangPRL2020}].

To analyze the RIXS spectra, our fitting model consists of the following three components. Elastic scattering and the bond-stretching phonon are described by Gaussian profiles. The paramagnon dominated background, at high energy, is mimicked using a damped harmonic oscillator function convoluted with the instrumental resolution [see Fig.~\ref{fig:fig1}(c)]. In Figs.~\ref{fig:fig2}(d) and \ref{fig:fig2}(e), subtraction of elastic and background responses isolates the bond-stretching phonon spectral weight.

In this fashion, phonon dispersion $E_{ph}(\mathbf{Q})$, lifetime effects and electron-phonon coupling are studied as a function of temperature and momentum across the charge ordering wave vector $\mathbf{Q}\rm_{CDW}$. Starting at high temperature (200~K), the phonon linewidth $W_{ph}$ is resolution limited, and the dimensionless parameter $W_{ph}/E_{ph}(\mathbf{Q})$ is essentially featureless across $\mathbf{Q}=\mathbf{Q}\rm_{CDW}$. At our base temperature (15~K), the phonon linewidth undergoes a weak but statistically significant enhancement beyond the resolution near $\mathbf{Q}\rm_{CDW}$, which indicates a decrease in the phonon lifetime. Combined with a softening of the phonon dispersion, $W_{ph}/E_{ph}(\mathbf{Q})$ displays a 20$\%$ increase across $\mathbf{Q}\rm_{CDW}$. Although current RIXS energy resolution barely resolves the phonon lifetime, the EPC can be analyzed through the softening of the dispersion and the RIXS phonon cross section.

Density functional theory (DFT) calculations~\cite{GiustinoNat2008,FalterPRB2006}, yield a bond-stretching phonon mode that disperses weakly downward from the zone center to the boundary [see gray dashed lines in Figs.~\ref{fig:fig3}(a) and \ref{fig:fig3}(c)]. The phonon softening observed around the charge ordering wave vector is not captured by the DFT formalism. By subtracting the bare phonon dispersion from the measured data (see Supplemental Material~\cite{SM_note} for details), we infer the softening magnitude [Fig.~\ref{fig:fig3}(e)]. The maximum value $\sim5$~meV at 15~K, is in good agreement with previous INS results on La-based compounds~\cite{TejsnerPRB2020,ReznikPhysC2012}.

\begin{figure}[ht!]
 	\begin{center}
 		\includegraphics[width=0.45\textwidth]{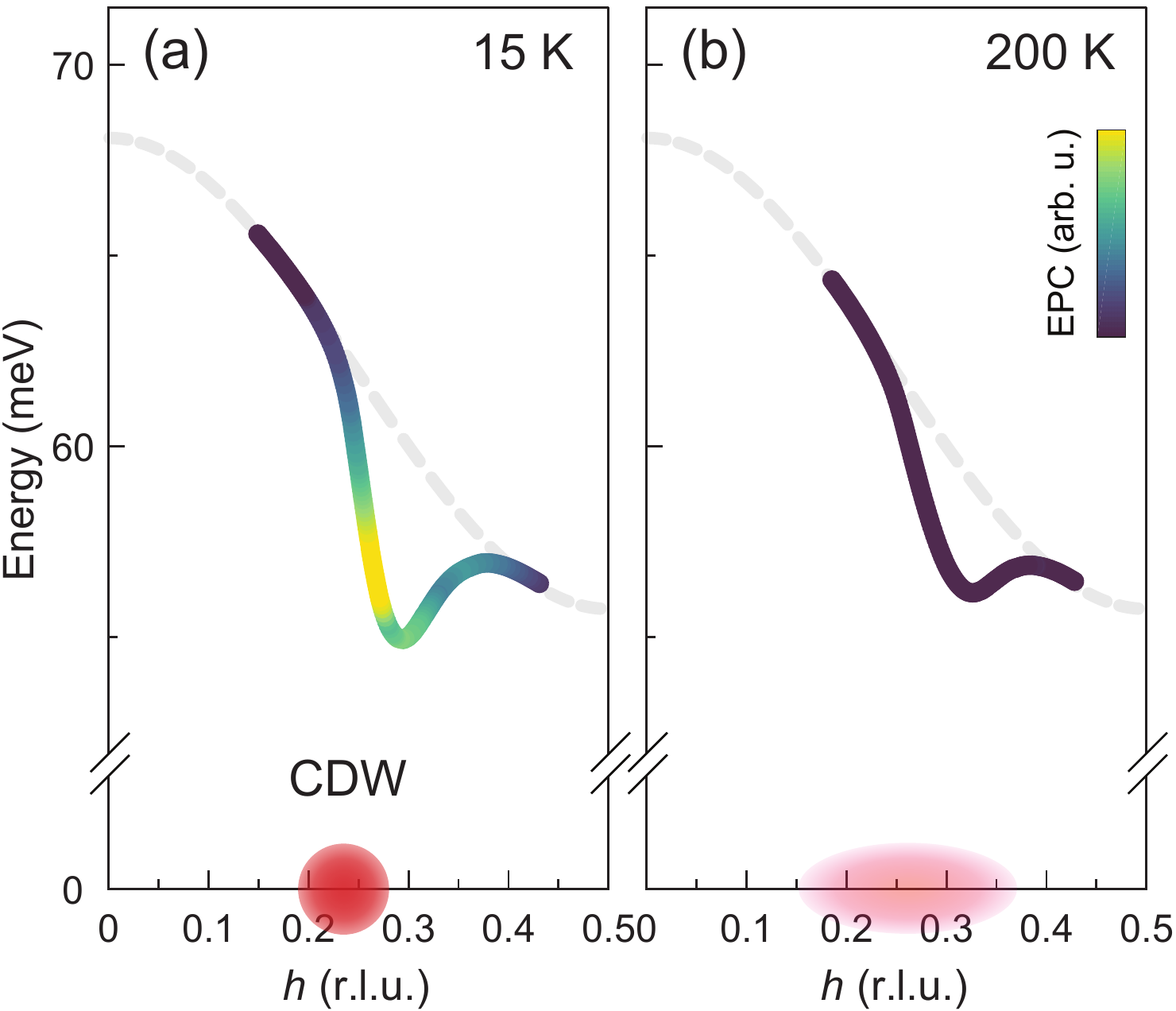}
 	\end{center}
 	\caption{
 	Schematic temperature evolution of the charge order and electron-phonon coupling through the bond-stretching mode. (a) Low- and (b) high-temperature charge ordering wave vector and soft bond-stretching phonon mode as observed in \EuLSCO. Red and pink area centered around the CDW ordering wave vector indicate schematically the CDW peak width. Solid and dashed lines represent the experimentally observed and DFT calculated bond-stretching phonon dispersions, respectively. Color code on the phonon dispersion indicates the observed electron-phonon coupling in addition to the expected $\sin^2(\pi h)$ dependence.
 	}	
	 	\label{fig:fig4}
 \end{figure}

In contrast to INS and IXS, the phonon intensities probed by RIXS are directly proportional to the electron-phonon coupling~\cite{AmentRMP2011,AmentEPL2011,BraicovichPRR2020,RossiPRL2019,DevereauxPRX2016,JohnstonPRB2010}. Above the LTT onset temperature ($T\rm_{LTT}\approx125$~K), the phonon integrated spectral weight increases monotonically with momentum $\mathbf{Q}=(h,0)$ [Figs.~\ref{fig:fig3}(b) and \ref{fig:fig3}(d)]. Upon cooling into the LTT phase, the phonon weight is enhanced in vicinity to $\mathbf{Q}\rm_{CDW}$ [Fig.~\ref{fig:fig3}(b)]. This effect is further enhanced when cooling to base temperature [Fig.~\ref{fig:fig3}(d)]. Our main finding is thus that electron-phonon coupling, linked to the charge ordering, is amplified within the LTT phase (see Fig.~\ref{fig:fig4}).

The phonon softening manifests over a broad momentum range ($h=0.24\mbox{-}0.4$) [Figs.~\ref{fig:fig3}(a) and \ref{fig:fig3}(c)]. This is in contrast to a conventional Kohn anomaly with a sharp phonon dispersion dip, and is thus incompatible with a CDW state arising from Fermi surface nesting~\cite{KohnPRL1959,ReznikPhysC2012,ZhuPNAS2015}. Alternatively, phonon anomalies in cuprates have been discussed in the context of collective stripe fluctuations~\cite{KaneshitaPRL2002,ParkPRB2014}. Phase excitations of stripes couple to the lattice and create a splitting of the in-phase and out-of-phase oscillation modes~\cite{KaneshitaPRL2002}. However, the phonon anomaly is expected~\cite{KaneshitaPRL2002} to exist within a narrow range of $\mathbf{Q}$ --- at odds with our experimental observations. Another notable effect of charge excitation is its Fano interference with intersecting optical phonons. Recent RIXS results on Bi-based cuprates Bi$_{2}$Sr$_{2}$CaCu$_{2}$O$_{8+\delta}$ and Bi$_{2}$Sr$_{2}$LaCuO$_{6+\delta}$ are interpreted in terms of a Fano resonance between dispersive charge excitations and optical phonon modes~\cite{LeeNP2020,ChaixNP2017,LiPNAS2020}. This is supported by a non-monotonically increasing phonon intensity versus momentum and an excitation energy softening that exceeds expectations for electron-phonon coupling. By contrast, the here reported softening magnitude in LESCO is consistent with typical phononic self-energy effects~\cite{ReznikPhysC2012,TejsnerPRB2020}. In addition, the momentum dependence of the phonon intensity shows marked differences. In LESCO, the enhanced RIXS intensity peaks around $\mathbf{Q}\rm_{CDW}$ (Fig.~\ref{fig:fig4}). By contrast, in Bi-based compounds, the intensity enhancement is not directly connected to the charge ordering wave vector. Electron-phonon coupling therefore appears as the most plausible mechanism underlying the observed phonon anomaly in LESCO. The different behaviors observed in LESCO and Bi-based compounds could be rooted in the charge ordering strength. Shorter correlation length found in Bi-based compounds makes quantum fluctuations correspondingly more relevant~\cite{LeeNP2020,LiPNAS2020}.

Although this phonon anomaly  is linked to the charge ordering phenomenon, there is no direct proportionality between charge order and phonon softening. For example, a pronounced phonon softening is observed already at 200~K in the presence of only weak charge ordering. Upon cooling the charge order grows by roughly two orders of magnitude whereas the phonon softening develops only marginally. The broad momentum width of the softening at all temperatures remains comparable to the short-range charge order peak above $T\rm_{LTT}$ [see Figs.~\ref{fig:fig3}(e) and \ref{fig:fig3}(f)]. All together, these results suggest a strong coupling between fluctuating stripes and the bond-stretching phonon.

Momentum dependence of the electron-phonon coupling in cuprates has been the subject of many theoretical considerations~\cite{AmentEPL2011,BraicovichPRR2020,RossiPRL2019,DevereauxPRX2016,JohnstonPRB2010}. For the copper-oxygen bond-stretching mode, a $\sin^2(\pi h)$ dependence of intensity for $\mathbf{Q} = (h,0)$ is expected in absence of charge order~\cite{BraicovichPRR2020,RossiPRL2019,DevereauxPRX2016,JohnstonPRB2010}. Recent RIXS studies, on different cuprate systems, have observed bond-stretching phonon intensities in good agreement with this theoretical modelling~\cite{RossiPRL2019,BraicovichPRR2020,LinPRL2020,LiPNAS2020}. It is also in accord with the intensities measured above $T\rm_{LTT}$ in LESCO [Figs.~\ref{fig:fig3}(b) and \ref{fig:fig3}(d)]. This linkage between RIXS phonon intensity and electron-phonon coupling suggests that upon cooling into the LTT phase, the EPC is enhanced around the CDW wave vector. This enhancement is further amplified upon cooling.

In La$_{2-x}$Sr$_x$CuO$_4$ $x=1/8$ (LSCO), the RIXS intensity of bond-stretching phonon remains structureless at low temperature (28~K)~\cite{LinPRL2020}. The lack of LTT lattice distortion in LSCO results in a mismatch between the copper-oxygen bond and the average stripe directions~\cite{KimuraPRB2000,ThampyPRB2014}. This in turn reduces the coupling between the fluctuating stripes and the lattice vibrations that involve the in-plane copper-oxygen bonds. Our comparison of the bond-stretching phonon across the LTT transition temperature in LESCO, indeed reveals that the EPC is enhanced at the charge ordering wave vector only in the LTT phase~[Fig.~\ref{fig:fig3}(b)]. As a result of the enhanced coupling between lattice and charge stripes, the charge order correlation length gradually increases inside the LTT phase~[inset of Fig.~\ref{fig:fig3}(f)]. The link between the lattice LTT transition, electron-phonon coupling, and charge correlations suggests that electron-lattice interactions promote the charge ordering process. 

The importance of EPC in cuprates has been underlined in many theoretical considerations~\cite{JohnstonPRB2010,DevereauxPRX2016,BraicovichPRR2020,BanerjeeComPhy2020}.
A phonon-based CDW mechanism can be enabled by strong correlations where the momentum space structure of the EPC is linked to the charge ordering wave vector~\cite{BanerjeeComPhy2020}. Although only the buckling mode was considered~\cite{BanerjeeComPhy2020}, the EPC of the bond-stretching mode is also known to be significant~\cite{ReznikNat2006,JohnstonPRB2010,DevereauxPRX2016,RossiPRL2019,BraicovichPRR2020,PengPRL2020}. Our experimental results indeed reveal that charge correlations spontaneously emerge at high temperature where the EPC is not favoring charge instability at any specific momentum. The low temperature EPC enhancement promotes the CDW formation by triggering a ``lock-in" of charge modulation at a specific ordering wave vector. This is similar to the conventional phonon mechanism in CDW-susceptible materials, where the $\mathbf{Q}$-dependent EPC induces a structure instability that has a compatible modulation with the charge density~\cite{MazinPRB2008,ZhuPNAS2015,BanerjeeComPhy2020}. In LESCO, the LTT distortion changes the in-plane copper-oxygen bond lengths with the same symmetry breaking tendency as the stripes. The ionic modulation could possibly enhance the coupling to the electronic degree of freedom and stabilize the charge stripes. Future improvements of the RIXS energy resolution would enable studies of energetically lower lying phonons --- for example the buckling mode~\cite{LiPNAS2020}. Such studies would map out all the charge ordering relevant phonon modes.

In summary, we use RIXS to study self-energy effects and electron-phonon coupling of bond-stretching phonon anomaly associated with charge-density-wave order in \EuLSCO. A substantial enhancement of the electron-phonon coupling is observed near the charge ordering wave vector upon entering the low-temperature tetragonal phase. A tangible causal chain is that the LTT lattice structure enables a momentum dependent electron-phonon coupling that in turn triggers a lock-in of the charge-stripe ordering wave vector and long-range charge correlations. Our results highlight the importance of electron-phonon interaction for the long-range CDW order in the cuprates. Since CDW is a competing order, the enhanced electron-phonon coupling on the bond-stretching mode, most likely, plays an effectively antagonistic role towards superconductivity.

\begin{acknowledgments}
\section{Acknowledgments}
Q.W. and J.C. thank Zhenglu~Li, Steven~G.~Louie and Yingying Peng for insightful discussions. Q.W., K.v.A., M.H., T.S., and J.C. acknowledge support by the Swiss National Science Foundation through Grant Numbers BSSGI0-155873, 200021-188564. Y.S. is funded by the Swedish Research Council (VR) with a Starting Grant (Dnr. 2017-05078). We acknowledge Diamond Light Source for time on Beamline I21 under Proposal MM24481. Q.W. and K.v.A. contributed equally.
\end{acknowledgments}

\end{document}